\documentclass{aa}                     
\usepackage{epsfig}
\usepackage{graphics}
\usepackage{float}
\usepackage{amsmath}
\usepackage{multirow}
\usepackage{longtable}
\usepackage{rotate}
\usepackage{array}
\usepackage{subfigure}
\def\kms{\ifmmode{\rm km\,s^{-1}}\else\hbox{$\rm km\,s^{-1}$}\fi}

\setlongtables

\begin{document}

\title{Spectroscopic binaries with elliptical orbits}

\author{L.B.Lucy}
\offprints{L.B.Lucy}

\institute{Astrophysics Group, Blackett Laboratory, Imperial College 
London, Prince Consort Road, London SW7 2AZ, UK}
\date{Received ; Accepted }

\abstract{The radial velocity curves of many spectroscopic binaries (SBs) 
are perturbed
by gas streams or proximity effects.
For SBs with circular orbits,  
these perturbations can give rise to spurious orbital eccentricities of
high statistical significance. But tests to identify such
anomalous orbits can be constructed since perturbed velocity
curves are in general no longer Keplerian. The derived tests are 
applied both to synthetic and to observed 
velocity curves.
   }

\authorrunning{Lucy}
\titlerunning{Elliptical orbits}
\maketitle
%________________________________________________________________

\section{Introduction}

Observers of spectroscopic binaries and exoplanets commonly 
test the statistical
significance of orbital eccentricities if $e \la 0.1$. The widely adopted
methodology (Lucy \& Sweeney 1971, 1973) 
is to test the null hypothesis $H_{0}$ that $e = 0$. If $H_{0}$ is
rejected at the 5 per cent level, the elliptical orbit is accepted.
If not, the circular orbit is preferred.   

	Although recommending and implementing this methodology, 
Lucy \& Sweeney (1971, hereafter LS) noted that 
some $e$'s thus accepted 
are spurious, being due to absorption line distortions by gas
streams or proximity effects.  
Accordingly, they suggested that, if $ e > 0$ is essential for a subsequent 
investigation, then the 1 per cent level of 
significance could be adopted. Nevertheless, they noted that the well known
short-period eclipsing binary U Cephei, for which photometric evidence
indicates $e = 0$, is still assigned an elliptical orbit at the 1 and even 
the 0.1 per cent level of significance.

	Given that changing the level of significance in the LS-test
is not a secure method of identifying a sample of SBs free from 
spurious $e$'s, a different approach is required to detect
perturbations. The expectation that a perturbed velocity curve will in
general not be strictly Keplerian is the basis for the additional tests
developed in this paper.  

	This investigation is prompted by the work of Skuljan et al.
(2004). Exploiting the high-precision radial velocities attainable with
a modern spectrograph, they report new orbital elements for the single-lined
SB $\zeta$ TrA derived from measurements with standard
error $\sigma = 0.014$ km s$^{-1}$, a vast improvement on the previous
orbit by Spencer Jones (1928) for which $\sigma = 1.4$ km s$^{-1}$ -
see entry no. 445 in Table I of LS. With this jump in precision, it is
no great surprise that a statistically insignificant $e$ of
$0.06$ is now replaced by a highly significant value of
$0.0140 \pm 0.0002$. 
Moreover, the other elements are correspondingly improved in precision.
In particular, $\sigma_{\omega}$, the error on the longitude of periastron
$\omega$
is reduced from $63 \degr$ to a mere $0 \degr .9$.

	With this remarkable reduction in $\sigma_{\omega}$,
the time interval required for the detection of apsidal motion for
this and similar SBs is
greatly reduced, raising the prospect of greatly increasing the
data set for this classic test of stellar structure . But before making
second epoch observations,
an investigator should check that $e$, no matter how significant
according to the LS-test,
passes the further tests developed in this paper.

\section{Harmonic analysis}

Nowadays, orbital elements are almost exclusively determined
by least-squares fitting; and the LS-test
as well as the tests developed herein are specific to such 
solutions.
Nevertheless, these new tests are more readily understood
in the context of determining the elements from a harmonic analysis of
the velocity curve (Wilsing 1893, Russell 1902, Monet 1979).    
\subsection{Fourier expansion}
The Keplerian velocity $V$ of a component of an SB with orbital
period $P$ is
\begin{equation}
 V = \gamma + K [\, cos(\nu + \omega) + e \, cos \, \omega ]
\end{equation}
where $\gamma$ is the radial velocity of the barycentre, $K$ is
the semi-amplitude, and $\nu(t)$ is the true anomaly. Since the 
motion is periodic, $V(t)$ can be expanded as a Fourier series,
\begin{equation}
 V = \frac{1}{2} C_{0} + \sum_{j = 1}^{\infty} \left [ \, C_{j} \, cos \, jL +
                                            S_{j} \, sin \, jL \right ]
\end{equation}
where $L = 2 \pi (t-T_{0})/P$ is the mean longitude.

   For given orbital elements
$P, T_{0}, \gamma, K, ecos \omega,  esin \omega$,  the Fourier coefficients
$C_{j}$ and $S_{j}$ are derivable from Eq. (1)
by numerical integration of the Fourier integrals.
Alternatively, for $e \la 0.5$, accurate analytic formulae can be
obtained 
by expanding in powers of $e$. Such formulae 
are given in the Appendix for the coefficients of the first four harmonics. 
\subsection{Estimating orbital elements}
Given numerous measurements of a component's radial velocity,
the orbital elements may be estimated as follows. First, $P$ is
determined by minimizing the scatter about the mean velocity curve. Then,
with a trial value of $T_{0}$,
the coefficients $C_{j}$ and $S_{j}$ are estimated by approximating their
Fourier integrals by summations over the observed velocities. The final 
step is to derive the orbital elements {\em algebraically} by equating these
estimates, $\tilde{C}_{j}$ and $\tilde{S}_{j}$, to their Keplerian
counterparts.

	With only the leading term retained from the expressions in
Appendix A, we have $C_{0} = 2 \gamma$, $C_{1} = K$, $S_{1} = 0$,
$C_{2} = Ke \, cos \omega$, and $S_{2} = Ke \, sin \omega$. Accordingly, if
$T_{0}$ is adjusted so that $\tilde{S}_{1} = 0$, then
$\gamma = \tilde{C}_{0}/2$,
$K = \tilde{C}_{1}$,
$e \, cos \omega =  \tilde{C}_{2}/K$, and
$e \, sin \omega =  \tilde{S}_{2}/K$.
These elements can be improved iteratively to allow for higher
order terms
in $e$ in the formulae for  $C_{1,2}$ and $S_{1,2}$.
\subsection{Perturbed velocity curves}
A perturbation due to gas streams or proximity
effects will 
result in estimates $\tilde{C}_{j}$ and $\tilde{S}_{j}$ that differ
systematically from their exact Keplerian values. But such differences
in  $\tilde{C}_{1,2}$ and $\tilde{S}_{1,2}$
do not cause the above 
solution procedure to fail, and so the investigator remains unaware both of
the perturbation
and of the resulting systematic errors in the elements. 

	To detect a perturbation, the analysis
must be extended to the third harmonic, for which the leading terms
are $C_{3} = 9/8 K e^{2} cos 2 \omega$ and
$S_{3} = 9/8 K e^{2} sin 2 \omega$. Evidently, $C_{3}$ and $S_{3}$ are
predicted by the complete set of elements derived above from 
$\tilde{C}_{1,2}$ and $\tilde{S}_{1,2}$, leaving no adjustable parameters
to fit $\tilde{C}_{3}$ and $\tilde{S}_{3}$. Accordingly,
if the velocity curve is perturbed, the vector
($\tilde{C}_{3}, \tilde{S}_{3}$) will in general differ from the prediction
($C_{3}, S_{3}$). This is the basis for the test developed in Sect. 3.2 
\section{A hierarchy of tests}
In this section, tests are developed for separating
SBs with perturbed velocity curves from those consistent with pure
Keplerian motion. Each test compares the goodness-of-fits for a pair of
models, where one of the pair is derivable from the other by imposing
constraints on its parameters.

	The goodness-of-fit criterion for model $k$ is the sum of weighted
squared residuals
\begin{equation}
 R_{k} = \sum_{n} w_{n} (v_{n} - V_{n})^{2}
\end{equation}
where $v_{n}$ is the observed radial velocity with relative weight $w_{n}$,
and $V_{n}$ is the predicted velocity.

	This sequence of tests starts with the LS-test for 
statistically significant $e$'s. For completeness and
clarity, this test is briefly recapitulated here.   
\subsection{Detection of ellipticity}
Given the data set ($v_{n},w_{n},t_{n}$),
we first calculate the least-squares circular orbit ($k = 1$), thus
determining the elements $P, T_{0}, \gamma$ and $K$. Next, 
the least-squares elliptical orbit ($k = 2$) is computed, thus determining
the elements   
$P, T_{0}, \gamma,K, e \, cos \omega, e \, sin \omega$. Because of the
two additional parameters, the elliptical orbit is always a better fit to
the data, so that $R_{2} < R_{1}$. The question is whether the reduction 
justifies rejecting the circular orbit. To assess this,
LS treated the problem as one of testing multivariate linear hypotheses.
In this case, the null hypothesis of a circular orbit is obtained by
adopting the following linear hypotheses 
\begin{equation}
 e \, cos \omega = 0  \; ,  \;\;\;  e \, sin \omega = 0
\end{equation}
with respect to {\em two} of the orbital elements.

	Applying standard statistical theory, LS found
that the probability that measurement errors of an SB with $e = 0$
could result in a least-squares elliptical orbit with $R \le R_{2}$ is 
\begin{equation}
 p_{1} = \left ( \frac{R_{2}}{R_{1}} \right )^{\beta_{2}} 
\end{equation}
Here $\beta_{2} = 1/2(N-M_{2})$, where $N$ is the number of measured
velocities, and
$M_{2}$ is the number of elements for the elliptical orbit - i.e., for the
unconstrained model.

	The probability $p_{1}$ determines the outcome of the first decision
we make about this SB: the circular orbit is accepted if $p_{1} > 0.05$,
and the elliptical orbit if $p_{1} < 0.05$.   

	If the $v_{n}$ were subject only to measurement
errors, this decision would conclude the investigation. Thus, if
$p_{1} < 0.05$, the reality of $e$ would be accepted.
But now we question the reality of such statistically significant
$e$'s.
\subsection{Detection of perturbation}
In Sect. 2.3, we concluded that a perturbation will in general cause
a displacement between the vectors
($\tilde{C}_{3}, \tilde{S}_{3}$) and ($C_{3}, S_{3}$). In the context of a
least-squares solution, this implies that $v_{n} - V_{n}$, the residuals
remaining
after fitting an elliptical orbit, will still contain a 3rd harmonic
component.   

	To look for this, consider the following two-parameter model
\begin{equation}
 V = V_{ell} + \Delta C_{j} cos jL + \Delta S_{j} sin jL
\end{equation}
Here $V_{ell}$ is the previously-determined elliptical orbit,
and the additional terms are
an adjustable $j$th harmonic, with $j = 3$ unless otherwise
specified.
The least-squares solution for this model ($k = 3$) determines the
vector $\vec{\Delta_{j}} = (\Delta C_{j}, \Delta S_{j})$, with the best-fit  
subsequently written as
$\vec{\delta_{j}} = (\delta C_{j}, \delta S_{j})$.

	Given its adjustable  
parameters, model 3 improves the fit, so that $R_{3} < R_{2}$.
But is this reduction large enough for the null hypothesis of an
unperturbed elliptical to be rejected? Because the null hypothesis
is obtained by imposing {\em two} constraints, 
\begin{equation}
      \Delta C_{j} = 0 \;,  \;\;\;   \Delta S_{j} =0
\end{equation}
the required test is
identical to that of LS. Thus, the probability that
measurement errors of the unperturbed elliptical orbit could give 
$R \le R_{3}$ for model 3 is  
\begin{equation}
 p_{2} = \left ( \frac{R_{3}}{R_{2}} \right )^{\beta_{3}} 
\end{equation}
with $\beta_{3} = 1/2(N-M_{3})$, where $M_{3} = 2$ is the number of
parameters for the unconstrained model.

	The probability $p_{2}$ determines the outcome of the
first decision
made about the {\em reality} of $e$.
If $p_{2} > 0.05$, the least-squares displacement vector
$\vec{\delta_{3}}$
is consistent with measurement errors. There is then no evidence for a  
perturbation, and so $e$ passes this first test of reality.
This success implies that
the amplitudes and phases of the first three harmonics of the velocity 
curve are consistent with Keplerian motion. But if $p_{2} < 0.05$, a
perturbation
is detected, and so $e$ is suspect. Whether then to accept or reject
$e$ is discussed further in Sect. 3.4. 
\subsection{Detection of Keplerian harmonic}
Fitting Eq.(6) to the velocity curve tells us that its 
3rd harmonic can be written as 
$\vec{H_{3}} =\vec{K_{3}} + \vec{\delta_{3}}$, where 
$\vec{K_{3}} = (C_{3}, S_{3})$.

	Independently of the significance of $\vec{\delta_{3}}$
according to the $p_{2}$-test, we can ask if $\vec{K_{3}}$
is detected. This harmonic is regarded as detected if the hypotheses
that $\vec{K_{3}}= \vec{0}$ and 
$\vec{H_{3}}= \vec{0}$ are both rejected. The second hypothesis is
necessary in order to reject solutions that imply a fortuitous near 
cancellation of $\vec{\delta_{3}}$ and $\vec{K_{3}}$.

     To test the first hypothesis, $\vec{K_{3}}$
is subtracted from the solution
of Sect. 3.2 in order to see if it is essential 
to that fit's success.
This subtraction is effected by again using Eq. (6) but now with
constraints 
\begin{equation}
   \Delta C_{j} = \delta C_{j} - C_{j} \; , \;\;\
             \Delta S_{j} = \delta S_{j} - S_{j} 
\end{equation}

   The model thus constrained is {\em not} a physically realistic model
of the
velocity curve. Instead, it defines a null hypothesis set up as a
`straw man' purely for testing the significance of $\vec{K_{j}}$.   
Accordingly, if the `straw man' survives the test, we do not
accept the null hypothesis but merely conclude that $\vec{K_{j}}$ 
is too small to be detected. 

	With $\vec{\Delta_{j}}$ fixed by Eq. (9),
this model ($k = 4$) has no adjustable parameters, so that $R_{4}$
is derived arithmetically. Moreover, since Eq. (9) represents
a displacement from the least-squares solution, $R_{4} > R_{3}$. 
But is this increase large enough to reject the null hypothesis? 
Because again {\em two} constraints have been imposed, the
test is formally identical to that of Sect. 3.2. Thus, on the null
hypothesis that model $k = 4$ is the true model,
\begin{equation}
 q_{1} = \left ( \frac{R_{3}}{R_{4}} \right )^{\beta_{3}} 
\end{equation}
is the probability
that measurement errors could give $R \le R_{3}$ when
model 3 is fitted to the data.

     To test the second hypothesis, $\vec{H_{3}}$
is subtracted from the least-squares solution by again using Eq. (6) but now
with constraints 
\begin{equation}
   \Delta C_{j} = - C_{j} \; , \;\;\     \Delta S_{j} = - S_{j} 
\end{equation}
With $\vec{\Delta_{j}}$ thus fixed,
this model ($k = 5$) also has no adjustable parameters, so that $R_{5}$
is derived arithmetically. Moreover, since Eq. (11) represents
a displacement from the least-squares solution, $R_{5} > R_{3}$. 
But is this increase large enough to reject the hypothesis? 
Because again {\em two} constraints have been imposed, the
test is again formally identical to that of Sect. 3.2. Thus, on the
hypothesis that model 5 is the true model,
\begin{equation}
 q_{2} = \left ( \frac{R_{3}}{R_{5}} \right )^{\beta_{3}} 
\end{equation}
is the probability
that measurement errors could give $R \le R_{3}$ when
model 3 is fitted to the data.

   Given the outcomes of these two tests, we now define
\begin{equation}
 p_{3} = \max (q_{1}, q_{2}) 
\end{equation}
which is such that $p_{3} < 0.05$ implies $q_{1} < 0.05$ {\em and} 
$q_{2} < 0.05$. 
Accordingly, if
$p_{3} < 0.05$, $\vec{K_{3}}$ is detected. But
if $p_{3} > 0.05$, $\vec{K_{3}}$ is not regarded as detected, even if 
$q_{1} < 0.05$.
\subsection{Decisions}
Following the $p_{1}$-test's detection of a significant
$e$, its reality is subject to the $p_{2}$- and $p_{3}$-tests.
There are four possible outcomes: 
\subsubsection{$p_{2} > 0.05 \,; \;\; p_{3} < 0.05$} 
The eccentricity of an SB in this domain of
($p_{2},p_{3}$)-space 
is powerfully supported. Not only are the amplitudes and phases
of the first three Keplerian harmonics consistent with the data, with no
evidence of
perturbation, but $\vec{K_{3}}$ is detected. Thus the
interpretation of the velocity curve's 2nd harmonic $\vec{H_{2}}$
as due to orbital
eccentricity is decisively confirmed. 
Follow-up investigations that require $e > 0$ can be carried out with
complete confidence.
\subsubsection{$p_{2} < 0.05 \,; \;\; p_{3} < 0.05$} 
The eccentricity of an SB in this domain  
has some support. Despite $\vec{\delta_{3}}$ being significant,
the detection of $\vec{K_{3}}$ indicates an elliptical
orbit. Nevertheless, the elements are
uncertain because of the perturbation.
Follow-up investigations that require $e > 0$
are therefore risky because of the unknown systematic errors of $e$ and
$\omega$.
\subsubsection{$p_{2} > 0.05 \,; \;\; p_{3} > 0.05$} 
The eccentricity of an SB in this domain  
is not strongly supported. Although $\vec{\delta_{3}}$
is not significant, the failure to detect $\vec{K_{3}}$ implies that
$\vec{H_{2}}$ may still be due to a 2nd harmonic perturbation - see
Sect. 4.1. Accordingly,
follow-up investigations that require $e > 0$ are subject to risk.
\subsubsection{$p_{2} < 0.05 \,; \;\; p_{3} > 0.05$} 
The eccentricity of an SB in this domain of
($p_{2},p_{3}$)-space 
is in doubt. The amplitudes and phases
of the first three Keplerian harmonics are inconsistent with the data,
thus indicating a perturbed velocity curve. Moreover, the non-detection
of $\vec{K_{3}}$
undermines the interpretation of $\vec{H_{2}}$
as due to an elliptical orbit. Follow-up investigations that require $e > 0$
are inadvisable.
\subsection{Comments}
The following comments are intended to clarify the above
tests:\\
\indent
	a) If a circular orbit is not rejected by the $p_{1}$-test, its
acceptance is recommended.

	Some spectroscopists assume that the motivation for thus
setting $e = 0$ is a preference for simplicity; others
believe that this decision is based only on statistical arguments.
But LS (Sect. II) also stressed the {\em physical} argument `that the
effect of tidal friction is to diminish the eccentricity of a binary
system.' Accordingly, when they discovered (LS, Sect. V) a high frequency of
circular orbits for long-period SBs with giant components, they
attributed such orbits to `energy dissipation in the deep convective
envelopes of the giant component.' A further discussion of the
{\em a posteriori} evidence for the efficacy of tidal dissipation 
is given by Lucy \& Sweeney (1973).

	When an observer finds that $e$ is not significant,
he can claim, as did LS, to have discovered that a mechanism causing
a secular decrease
of $e$ has been operating. If so, it is unlikely that
the secular decrease has been caught at the special epoch when 
$e \approx E(\hat{e})$, the expected value due to measurement
errors when $e = 0$ - Eq. (18) in LS.
Accordingly, if $e$ is not significant, we expect that typically
$e \ll E(\hat{e})$, and so the best option is to set $e=0$. 

	Of course, an observer may choose not to claim this discovery and
so publish the elliptical orbit. But the resulting contamination
of catalogues with $e$'s that can be attributed to measurement
errors is regrettable and misleading. A theorist interprets a small
but non-zero $e$ as indicating a modest number of $e$-foldings in the
secular decrease of $e$. But if $e$ is not 
significant, the data is consistent with an infinite number $e$-foldings.

	The statistical and physical arguments for setting $e = 0$ when
$p_{1} > 0.05$ were already compelling in the early 1970's, and they are 
even more so now.\\
\indent
	b)  If a circular orbit is accepted, its
reality is not doubted. A perturbation that converts an elliptical orbit
into an apparently circular one is possible but improbable.\\ 
\indent
	c) Because $e$ is non-negative, errors in $e$ are non-gaussian when
$e \simeq 0$. In consequence, the estimate of $e$ derived from 
$e cos \omega$ and $e sin \omega$ has a positive bias, and this 
can mislead the investigator into concluding that $e$ is significant.
To avoid such misjudgements, LS applied the statistical theory of
hypothesis testing.

	In this paper, the detectability of small, non-negative Fourier
amplitudes is an issue. Again, the hypothesis-testing approach 
is preferable to comparing (biased) amplitudes to their standard errors.\\
\indent
	d) Although $p_{1} < 0.05$ is interpreted in Sect. 3.1 as the
detection of a significant $e$, it could more accurately be
described as the detection of a significant $\vec{H_{2}}$ - see Sect. 2.2.
Since perturbations by gas streams or proximity effects can introduce a
2nd harmonic even when $e=0$, the detection of $\vec{H_{2}}$ cannot
of itself be decisive evidence for an elliptical orbit.\\ 
\indent
	e) The statistical tests of Sects. 3.1-3.3 assume
normally-distributed measurement errors and models that are
linear in the adjustable parameters. The required linearity holds rigorously
for the $p_{2}$- and $p_{3}$-tests - see Eq. (6) - but not for the
$p_{1}$-test. 
For this latter test, Monte Carlo experiments (Lucy 1989) confirm its
validity.\\
\indent
	f) The adjustable 3rd harmonic in Eq. (6) is {\em not} a model
of the perturbation.
We assume only that the perturbation has period $P$ and has a 
Fourier expansion with non-zero amplitudes for one or more of the first
three harmonics.
In particular, even though we seek evidence of a perturbation via the third
harmonic,
the efficacy of the $p_{2}$-test does not require the perturbation to
contain this harmonic.

	To illustrate this point, let us consider an error-free
velocity curve subject to the 1st harmonic perturbation
$ \epsilon  K cos \, L$. The resulting modified elements,
denoted by primes, are derived using the leading-terms formulae from
Sect. 2.2.  
Since only the cosine term of the 1st harmonic is perturbed, we have
$\tilde{C_{1}} = (1 + \epsilon) C_{1} =  (1 + \epsilon) K$, with all other
estimated coefficients having
their exact values. Accordingly, we find $\gamma^{\prime} = \gamma$,
$K^{\prime} = K(1 + \epsilon) $,  $\omega^{\prime} = \omega$, and
 $e^{\prime} = e/ (1 + \epsilon)$.  
On the basis of these elements, we predict 
$C^{\prime}_{3} = C_{3}/(1 + \epsilon)$ and similarly for $S^{\prime }_{3}$.
Thus, as anticipated in Sect. 2.3, there is a discrepancy between the
predicted vector
$(C^{\prime}_{3}, S^{\prime}_{3}) = (C_{3}, S_{3})/(1 + \epsilon)$
and its estimate
$(\tilde{C_{3}}, \tilde{S_{3}}) = (C_{3}, S_{3})$.\\
\indent
	g) When the $p_{2}$-test detects a perturbation, the orbital
elements are subject to unkown systematic errors. No statistical procedure
can then extract unbiased elements from the velocity curve.
Additional data or a physical model of the perturbation is required to
derive more reliable elements.\\ 
\indent
	h) Eq. (13) in LS defines a quantity  
\begin{equation}
 \mu = \frac{\sigma}{K} \, \sqrt{\frac{2}{N}}
\end{equation}
which is a dimensionless measure of
a velocity curve's precision. In terms of $\mu$, $e$
should be detectable at the 5 per cent level if
\begin{equation}
 e > 2.45 \mu
\end{equation}
- see Eq. (22) in LS.

	An essentially identical analysis allows criteria to be derived
for the detection of Keplerian harmonics in perturbation-free orbits
with $e^{2} \ll 1$.
Thus, the 5 per cent criteria are
\begin{equation}
 e^{2} > 2.18 \mu
\end{equation}
for $\vec{K_{3}}$, and
\begin{equation}
 e^{3} > 1.84 \mu
\end{equation}
for $\vec{K_{4}}$. Generally, $\vec{K_{j}}$ should be detectable at the
$100p$ per cent level of significance if 
\begin{equation}
 e^{j-1} > \frac{\sqrt{-2 \ell n p}}{a_{j}} \, \mu
\end{equation}
where $a_{j}$ is given by Eq. (A.11).

	The inequality given by Eq. (16) defines the domain in
$(e, \mu)$-space within which $e$ can be confirmed
by the $p_{3}$-test. Ideally, an observing campaign should be designed to
obtain a velocity curve that falls in this domain.\\
\indent
	i) The $p_{2}$-test assumes that the perturbation has period $P$.
But some causes of perturbation violate this assumption. One example is
pulsations, another is spots,
and yet another is a third-body perturbation.
In general, such effects increase the residuals and hence the deduced
measurement error $\sigma$ and
are not detected by $p_{2}$.\\
\indent
	j) The $p_{2}$-test relies entirely
on radial velocities for detecting a perturbation. Phase-dependent line
strengths and line shapes may provide independent evidence of perturbation.
Such evidence may also resolve the ambiguity noted in Sect. 3.4.3.\\   
\indent
	k) Following common practice, we have adopted the
5 per cent level of significance in stating the decisions reached from
the values of $p_{1,2,3}$. But the level can be varied.
In particular, the impact of
erroneous conclusions on subsequent investigations
should be considered.\\
\indent
	l) Many astronomers are now familiar with 
defining confidence intervals about a minimum-$\chi^{2}$ solution in terms
of an increment $\Delta \chi^{2}$ - e.g. Press et al. (1992). The 
relevant statistical theory is related to that from
which the tests of Sect.3 are derived.

	In the limit $N \rightarrow \infty$, the tests of Sect.3 simplify to
\begin{equation}
   \Delta \chi^{2} = -2 \, \ell n \,p 
\end{equation}
so that $p < 0.05$ corresponds to $\Delta \chi^{2} > 5.99$. Eq. (19)
reproduces the column for $\nu = 2$ in the table on p.692
of Press et al. (1992), except that their $p$ is the enclosed,  
and ours the excluded, probability. 

	Note that, in contrast to Eq(19), the tests of Sect.3 do {\em not}
require $N \rightarrow \infty$. Monte Carlo simulations confirm
that they remain valid even when $N$ only slightly exceeds $M$.   
\section{Monte Carlo experiments}
In this section, the tests of Sect. 3 are applied to synthetic velocity
curves. With the exact elements known, these experiments investigate the
reliability of the statistical tests.
\subsection{Perturbed circular orbit} 
Probably the commonest error for SBs is the assignment of a significant
$e$ to a binary with vanishingly small $e$ but 
subject to a 2nd harmonic perturbation. To simulate this case, 
synthetic data sets $(v_{n}, t_{n})$ are generated from the formula
\begin{equation}
 v_{n} =  V_{n} + \epsilon K\, cos (2L_{n} - \varpi)
                                                      +\sigma z_{n}
\end{equation}
Here the first term $V_{n}$ is the exact radial velocity at $t_{n}$ for a
circular orbit of semi-amplitude $K$; the second term is a
2nd harmonic perturbation of dimensionless amplitude $\epsilon$; and the
third term is a random gaussian measurement error with variance $\sigma^{2}$.

   A least-squares elliptical orbit for such a velocity curve 
will have a spurious $e \approx \epsilon$ with 
$\omega  \approx \varpi$.
By repeated sampling, we can investigate the success of
the $p_{2}$- and $p_{3}$-tests in detecting these spurious $e$'s
despite their possible significance according to the $p_{1}$-test.

   The first experiment is as follows. For fixed $\epsilon$, we generate
and analyse $10^{3}$ synthetic velocity curves with random
$\varpi$'s and uniformly-distributed $L_{n}$.
Each data set comprises $N = 50$ velocities $v_{n}$ with
$\sigma/K = 0.02$, so that $\mu = 4 \times 10^{-3}$.
Finally, the least-squares orbits assume $P$ fixed at its exact value.
From the $10^{3}$ analyses, the fractions $f_{1,2,3}$ with
$p_{1,2,3} < 0.05$ are computed, and this is then repeated with $\epsilon$
increasing from $0$ to $0.15$.

   The results are plotted in Fig.1.
The $f_{1}$-curve shows that significant
$e$'s increase sharply from $\approx 5$ per cent at $\epsilon = 0$ to  
100 per cent for $\epsilon \ga 0.02$. The 50 per cent detection rate occurs
close to the value $\epsilon_{1} = 2.45 \mu = 0.0098$ expected heuristically 
from Eq. (15) - the point P$_{1}$.
\begin{figure}
\vspace{8.2cm}
\includegraphics{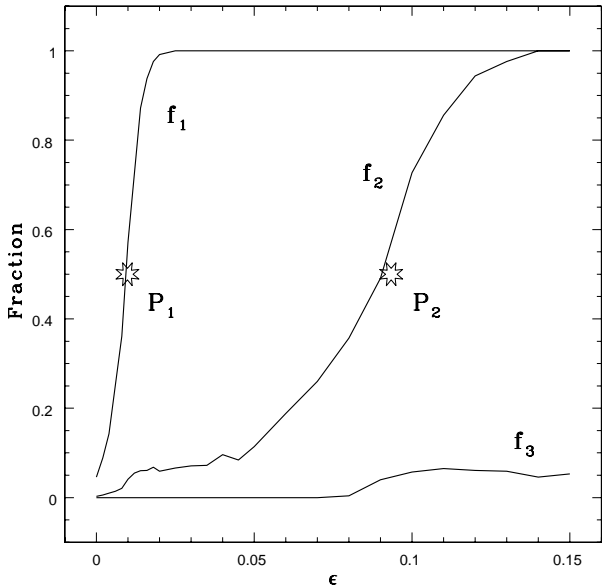}
\caption{Monte Carlo simulations. The fractions $f_{1,2,3}$ of synthetic
velocity curves with $p_{1,2,3} < 0.05$ are plotted against $\epsilon$,
the fractional amplitude of the 2nd harmonic perturbation in Eq.(20).
The starred points indicate the theoretical values of $\epsilon$ at which
typically
the spurious $e$ becomes significant ($P_{1}$) and the perturbation is
detected ($P_{2}$).}
\end{figure}
   The $f_{2}$-curve shows that for $\epsilon \la 0.06$ only a small fraction
of the significant $e$'s are detected as spurious by the 
$p_{2}$-test. But with further increase of $\epsilon$, the absence of
$\vec{K_{3}}$ in Eq. (20) becomes detectable, and the  
$f_{2}$-curve rises to give 100 per cent detections of $\vec{\delta_{3}}$ when
$\epsilon \ga 0.12$. The 50 per cent rate occurs close to the value
$\epsilon_{2} = 0.093$ similarly expected from Eq. (16) - the point P$_{2}$. 

   Finally, the $f_{3}$-curve shows that for $\epsilon \ga 0.09$, a small
fraction
($ \approx 0.06$) of spurious $e$'s are adjudged real by the $p_{3}$-test.

   This experiment demonstrates that, for perturbed circular orbits, the
tests of Sect. 3 perform as expected. But the results also emphasize
that if a significant $e$ is such that
\begin{equation}
       2.45 \mu \;\, \la \; e \; \la \; \sqrt{2.18 \mu}
\end{equation}
then we cannot decide if $e$ is real or spurious. In this interval, the
$p_{2}$-test cannot detect a spurious $e$, and the $p_{3}$-test fails to
confirm a real $e$.  
\subsection{Perturbed elliptical orbit} 
   In a second experiment, $e = 0$ in Eq. (20) is replaced by $e = 0.12$
with random $\omega$. Other details are unchanged.  

   From a sequence of simulations with $\epsilon$ increasing from 0 to 0.15,
we find that the $p_{2}$-test detects the perturbation with a reliability
of 50 and 90 per cent at $\epsilon \approx 0.035$ and $0.06$, respectively.
The $p_{3}$-test confirms the reality of $e$ with a reliabilty that drops from
$\approx 95$ per cent at $\epsilon = 0$ to  
$\approx 60$ per cent at $\epsilon = 0.15$. This drop when
$\epsilon \approx e$ is not surprising since the imposed perturbation in the
2nd harmonic then has an amplitude $\approx |\vec{K_{2}}|$.   
\subsection{Unperturbed elliptical orbits} 
   In a third experiment, the perturbation amplitude $\epsilon = 0$
and $e$ is now the varied parameter.

From a sequence of simulations with $e$ increasing from 0 to 0.13, we find,
as expected from Sect. 3.5 i), that the $p_{1}$-test has a $\ga 50$ per cent  
succes rate in detecting $e$ when $e \ga 2.45 \mu$, and the 
$p_{3}$-test a $\ga 50$ per cent succes rate in confirming the reality
of $e$ when $e^{2} \ga 2.18 \mu$. 
\section{Examples}
In this section, the tests developed in Sect. 3 are illustrated by  
applying them to SBs with published orbits.
\subsection{HD 45088} 
This SB is the subject of the first paper
(Griffin \& Emerson 1975) in Griffin's series on spectroscopic orbits from
photoelectric radial velocities. The authors report a high quality
orbit with $e = 0.150 \pm 0.004$ derived from $N = 54$ measured
velocities with standard error $\sigma = 1.0$
km s$^{-1}$. The eccentricity is clearly highly significant but,
given the relatively short period of $6 \fd 992$, one might suspect
that the orbit is circular and $e$ spurious.

	The first step is to apply the LS-test for the significance of $e$.
This and subsequent calculations are carried out with $P$
fixed at the value determined by the original authors, so that 
$M_{2} = 5$. The result of
the LS-test is $p_{1} = 5.6 \times 10^{-38}$, confirming the
overwhelming significance of $e$.

	The second step is to compare this elliptical orbit 
to the fit obtained with Eq. (6) when $j = 3$.
The least-squares
fit gives $\delta C_{3} = 0.28 \pm 0.20$ km s$^{-1}$ and 
$\delta S_{3} = 0.07 \pm 0.19$ km s$^{-1}$, suggesting little or no
evidence for $\vec{\delta_{3}}$
beyond what can be attributed
to measurement errors. This is confirmed by
Eq. (8), which gives $p_{2} = 0.35$. 
The amplitudes and phases of the first three
harmonics of the velocity curve of HD 45088 are therefore consistent with
unperturbed Keplerian motion.

	The elliptical orbit of HD 45088
is now subject to the test of Sect. 3.3. The 
3rd harmonic coefficients computed from the least-squares  
elliptical orbit 
are $C_{3} = -1.26$ km s$^{-1}$ and $S_{3} = 0.58$ km s$^{-1}$. Subtracting
this component according to
Eqs. (6) and (9), then computing $R_{4}$, we find that
$q_{1} = 1.2 \times 10^{-7}$. Then, similarly subtracting  $\vec{H_{3}}$
according to Eqs. (6) and (11), we find that $q_{2} = 3.9 \times 10^{-6}$.
Accordingly, $p_{3} = q_{2} = 3.9 \times 10^{-6}$ and so
there is a highly significant
detection of $\vec{K_{3}}$. This confirmation of ellipticity and the
earlier non-detection of perturbation place 
HD 45088 in the domain
of ($p_{2},p_{3}$)-space containing orbits with the most reliable
$e$'s.

	Having detected $\vec{K_{3}}$, we can extend the
search to $\vec{K_{4}}$. However, the predicted coefficients 
$C_{4} = -0.15$ km s$^{-1}$ and $S_{4} = -0.19$ km s$^{-1}$
are somewhat too small for detection.\\
\begin{figure}
\vspace{8.2cm}
\includegraphics{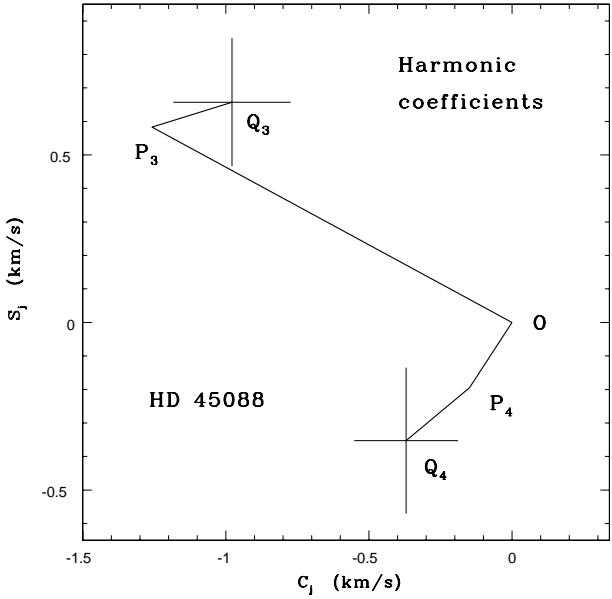}
\caption{Fourier coefficients for HD 45088. The 3rd and 4th harmonics
of the least-squares elliptical orbit are shown as the vectors
$\vec{OP_{3}}$
and $\vec{OP_{4}}$, respectively. The displacement vectors
 $\vec{\delta_{3,4}}$ obtained by fitting Eq.(6) are also plotted
together with their 1$\sigma$ error bars.
}
\end{figure}

	The calculations for the 3rd and 4th harmonics are illustrated
in Fig. 2. For the 3rd harmonic, we see that the components of 
$\vec{\delta_{3}} \equiv \vec{P_{3} Q_{3}}$
are comparable with
their standard
errors, in agreement with their non-significance according to $p_{2}$.
It is also evident that $\vec{K_{3}} \equiv \vec{OP_{3}}$ is the dominant
contribution to $\vec{OQ_{3}}$, so that its subtraction
significantly increases the residuals, leading to its highly significant
detection ($p_{3} \ll 0.05$).

	The corresponding vectors and error bars for the 4th harmonic are
also plotted in Fig. 2. These show that detection of this harmonic is
just beyond reach.

	The outcome of this investigation of HD 45088 is that 
it passes all tests that can be usefully carried out given the 
precision of the Griffin-Emerson velocity curve.
Needless to say, new data with $\mu$ reduced significantly below
the Griffin-Emerson value of $3.7 \times 10^{-3}$ might
reveal a perturbation that is currently undetectable. On the other hand,
if a perturbation were still not found, the confirmation of Keplerian motion
could be extended to the 4th and higher harmonics.

	For the above value of $\mu$, Eq. (15) predicts that $e > 0.009$
is required for the detection of $e$. This is consistent with $e = 0.15$
being overwhelmingly significant according to the $p_{1}$-test. Similarly,
Eq. (16) predicts that $e > 0.09$ is
required for the detection of $\vec{K_{3}}$, consistent with its 
highly significant detection according the $p_{2}$-test. Finally, Eq. (17)
predicts that $e > 0.19$ is required for the detection of $\vec{K_{4}}$,
consistent with its non-detection. 
\subsection{$\zeta$ Trianguli Australis} 
For this $12 \fd 976$ SB, Skuljan et al. (2004) report 225 radial
velocities obtained with a
high-resolution fibre-fed spectrograph. As noted in Sect. 1, the gain in
precision over the earlier orbit by Spencer Jones (1928) is a remarkable
two orders of magnitude. 

	Recomputing the orbit with $P$ fixed (their Table 6), we find
$p_{1} = 1.5 \times 10^{-148}$, in agreement with their claim of high
statistical significance despite the small eccentricity of 0.014.  

	Let us now investigate the reality of $e$ by applying the tests of
Sect. 3.2 and 3.3. The
first step is to fit Eq. (6) to their data. Switching the unit of velocity
to metres per s, we obtain
$\delta C_{3} = -3.3 \pm 1.3$ m s$^{-1}$ and 
$\delta S_{3} = -3.2 \pm 1.5$ m s$^{-1}$, which suggest a
significant perturbation. This is confirmed by finding
$p_{2} = 0.0063$. Accordingly, the null hypothesis of an 
unperturbed elliptical orbit is rejected.

	The next step is to seek confirmation of $\vec{K_{3}}$,
for which the predicted components are 
$C_{3} = -1.34$ m s$^{-1}$ and 
$S_{3} = 0.96$ m s$^{-1}$. Since these are comparable 
to the standard errors on the components of $\vec{\delta_{3}}$,
we immediately suspect that there is little
support for the detection of this harmonic. This is confirmed by
the $p_{3}$-test. Subtracting $\vec{K_{3}}$,
we find that $q_{1} = 0.47$. Similarly, subtracting $\vec{H_{3}}$,
we find that $q_{2} = 9.9 \times 10^{-4}$.
Accordingly, $p_{3} = q_{1} = 0.47$, and so $\vec{K_{3}}$ is not detected.

	These results place $\zeta$ TrA in the domain of ($p_{2},p_{3}$)-space
discussed in Sect. 3.4.4. This contains the least reliable category
of $e$'s. The dramatically improved precision has allowed
the discovery
of a very small amplitude yet highly significant $\vec{H_{2}}$.
But its
interpretion as due to orbital eccentricity is put in doubt by the
detection of a significant $\vec{\delta_{3}}$ {\em and} the non-detection of
$\vec{K_{3}}$.
The amplitude of $\vec{H_{2}} \sim 105$ m s$^{-1}$ is so small that
there is no difficulty in attributing it entirely to the detected
perturbation. 
\begin{figure}
\vspace{8.2cm}
\includegraphics{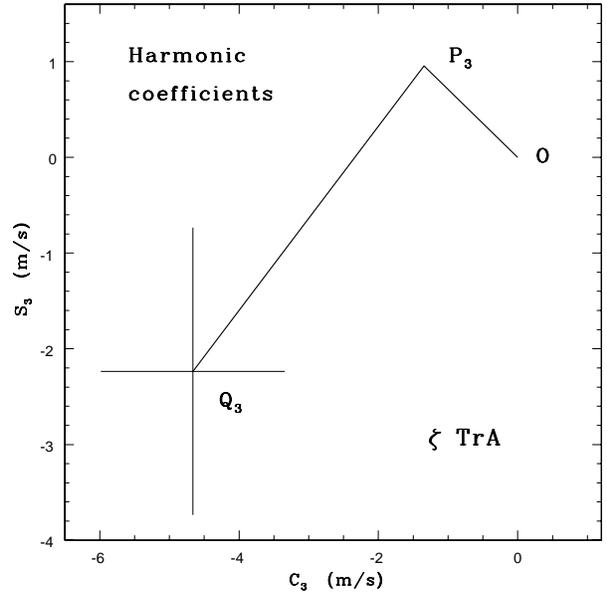}
\caption{Fourier coefficients for $\zeta$ TrA. The 3rd harmonic
of the least-squares elliptical orbit is shown as the vector
$\vec{OP_{3}}$.
The displacement vector $\vec{\delta_{3}}$ 
obtained by fitting Eq.(6) is also plotted
as $\vec{P_{3} Q_{3}}$ together with 1$\sigma$ error bars.
}
\end{figure}

	The above calculations are illustrated in Fig. 3. Comparison with
Fig. 2 for HD 45088 is illuminating. For HD 45088,
$\vec{OQ_{3}} \equiv \vec{H_{3}}$ is dominated by
$\vec{OP_{3}} \equiv \vec{K_{3}}$.
In contrast, for $\zeta$ TrA, $\vec{OQ_{3}}$ is 
dominated by $\vec{P_{3} Q_{3}} \equiv \vec{\delta_{3}}$.
This difference accounts for
detection of $\vec{K_{3}}$ for HD 45088 and its non-detection
for $\zeta$ TrA.     

	Despite the enormous statistical 
significance of the $e$ reported by Skuljan et al. (2004),
it fails to pass the tests of reality developed in Sect. 3.2 and 3.3. Thus
follow-up investigations predicated on the reality of $e$ carry substantial
risk of misleading or erroneous conclusions.
Nevertheless, just as 
$p_{1} > 0.05$ does not prove that an orbit is not eccentric but merely
that $e$ is not significant, so a finding that
$p_{2} < 0.05$ {\em and} $p_{3} > 0.05$ does not prove that $e$ is not
real but merely that this supposition does not find significant support
in the data. 
\section{Conclusion}
The aim of this paper has been to develop tests that allow
SBs with elliptical orbits of high reliability to be separated from those
with spurious $e$'s. The example of
HD 45088 shows that the literature contains orbits that are
not challenged by these tests. But 
$\zeta \, TrA$ shows that some published $e$'s
are, as expected, probably due to perturbations. Evidently, extensive further
recomputations 
would greatly clarify the reliability of the accumulated data on
SBs.
Indeed, categorizing SBs according to their location in 
the ($p_{1},p_{2},p_{3}$)-cube offers an objective treatment of orbit quality
that could replace or complement the subjective judgements
(e.g., $a,b,c, \ldots$) traditionally used by
compilers of
catalogues.
Note that the recently published 9th catalogue of SB orbits
(Pourbaix et al. 2004) compiles individual velocities, thus greatly
facilitating orbit recomputations.
  
	As to measuring apsidal motion with high precision velocity curves
(Sect. 1), observers should select SBs with
$p_{1} < 0.05$ and $p_{2} > 0.05$, and then aim for a new orbit
that satisfies the first and if possible also the second of the inequalities
given in Eqs. (16) and (17). By doing so, tests for the detection of
$\vec{K_{3}}$ 
and possibly also  $\vec{K_{4}}$ can be carried out. Such detections
would establish the reality of $e$ and thus justify
2nd epoch observations.

	In contrast, re-observing   
systems with $p_{1} > 0.05$ in the hope of discovering hitherto
undetected $e$'s runs the following risks: that $e$ remains
undetectable; that $e$ is significant but a perturbation is detected;  
that $e$ is significant and no perturbation is detected but $e$ is too
small for detection of $\vec{K_{3}}$. Evidently, this
(re-)observing strategy will have a much smaller discovery rate of orbits
in the highest category of reliability (Sect. 3.4.1).

	A long standing problem for orbits of SBs
is their non-uniform distribution of $\omega$, sometimes referred to as the
Barr
effect (see, e.g., Batten \& Ovenden 1968). This was present in
the orbits published by LS, both for those with and those without
significant $e$'s, and was attributed to perturbations. 
Accordingly, it will be of interest to re-examine the distribution of
$\omega$ for orbits with and without perturbations according to the
$p_{2}$-test. An obvious prediction is that orbits with $p_{2} > 0.05$ will
have a negligible
or much reduced Barr effect, and those with $p_{2} < 0.05$ will have the
effect enhanced.

	Given the typical precision of published velocity curves,
a programme of recomputation would investigate the effects of
perturbations with amplitudes $ \ga 0.1$km s$^{-1}$. But the work of
Skuljan et al. (2004) demonstrates that perturbations are detectable to much
lower amplitudes. In this regard, 
it would be interesting to observe eclipsing binaries
with light curves that imply $e cos \omega = 0$ to high precision.
Comparison with the spectroscopic values of $e cos \omega$ could then
reveal, or demonstrate the absence of, perturbations down to
$\sim 1$ m s$^{-1}$.

\appendix
\section{Fourier coefficients}

The Keplerian velocity of a binary component can be expanded in a Fourier
series in the mean anomaly $M$
(e.g., Plummer 1908). This is readily transformed into
the Fourier series in the mean longitude $L = M + \omega $ given in Eq. (2).
If we define $\alpha = K \, (1-e^{2}) \, cos \, \omega $ and
$\beta = K \sqrt{1-e^{2}} \, sin \, \omega $, the Fourier coefficients 
are
\begin{equation}
 C_{j} = \alpha \, c_{j} \, cos \, j \omega + \beta \, s_{j} \,
            sin \, j \omega
\end{equation}
and
\begin{equation}
 S_{j} = \alpha \, c_{j} \, sin \, j \omega - \beta \, s_{j} \,
            cos \,j \omega
\end{equation}
Expressions for the coefficients $c_{j}$ and $s_{j}$ in terms of Bessel
functions are given by Plummer (1908). Expansions of the latter as power
series in $e$ (e.g., Smart 1953, p.378) yields the following formulae,
accurate to $O(e^{7})$   
\begin{equation}
 c_{1} = 1 - \frac{1}{8} e^{2} + \frac{1}{192} e^{4}
                         - \frac{1}{9216} e^{6}
\end{equation}
\begin{equation}
 s_{1} = 1 - \frac{3}{8} e^{2} + \frac{5}{192} e^{4}
                         - \frac{7}{9216} e^{6}
\end{equation}
\begin{equation}
 c_{2} = e \left [1 - \frac{1}{3} e^{2} + \frac{1}{24} e^{4}
                     - \frac{1}{360} e^{6}  \right ]
\end{equation}
\begin{equation}
 s_{2} = e \left [1 - \frac{2}{3} e^{2} + \frac{1}{8} e^{4} -
                  \frac{1}{90} e^{6} \right ]
\end{equation}
\begin{equation}
 c_{3} = a_{3} \, e^{2} \left [ 1 - \frac{9}{16} e^{2} +
              \frac{81}{640} e^{4} \right ]
\end{equation}
\begin{equation}
 s_{3} = a_{3} \, e^{2} \left [ 1 - \frac{15}{16} e^{2} +
              \frac{189}{640} e^{4} \right ]
\end{equation}
\begin{equation}
 c_{4} = a_{4} \, e^{3} \left [ 1 - \frac{4}{5} e^{2}
                         + \frac{4}{15} e^{4}  \right ]
\end{equation}
\begin{equation}
 s_{4} =  a_{4} \, e^{3} \left [ 1 - \frac{6}{5} e^{2} 
                        +  \frac{8}{5} e^{4}  \right ]
\end{equation}
where
\begin{equation}
 a_{j} =  \frac{1}{(j-1)!} \, \left(\frac{j}{2} \right)^{j-1}
\end{equation}

	Values of $C_{j}$ and $S_{j}$ obtained with these
expansions have been checked against values obtained by numerical
integration of the Fourier integrals.

\end{document}